\documentclass[a4paper]{jpconf}
\usepackage{graphicx}
\usepackage{pifont}
\begin{document}
\title{Chemical complexity in NGC\,1068}

\author{R. Aladro$^1$, S. Viti$^1$, D. Riquelme$^2$, S. Mart\'in$^3$, R. Mauersberger$^4$, J. Mart\'in-Pintado$^5$, E. Bayet$^6$}

\address{$^1$ Department of Physics \&  Astronomy, University College London, Gower Street, London WC1E 6BT, UK.}
\address{$^2$ Instituto de Radioastronom\'ia Milim\'etrica, Avda. Divina Pastora, 7, Local 20, E-18012 Granada, Spain.}
\address{$^3$ European Southern Observatory, Avda. Alonso de C\'ordova 3107, Vitacura, Casilla 19001, Santiago 19, Chile.}
\address{$^4$ Joint ALMA Observatory,  Avda. Alonso de C\'ordova 3107, Vitacura, Santiago, Chile.}
\address{$^5$ Centro de Astrobiolog\'ia (CSIC-INTA), Ctra. de Torrej\'on Ajalvir km 4, E-28850 Torrej\'on de Ardoz, Madrid, Spain.}
\address{$^6$ Sub-Department of Astrophysics, University of Oxford, Denys Wilkinson Building, Keble Road, Oxford OX1 3RH, UK.}

\ead{r.aladro@ucl.ac.uk}

\begin{abstract}

We aimed to study the chemistry of the circumnuclear molecular gas
of NGC\,1068, and to compare it with those of the starburst
galaxies M\,82 and NGC\,253. Using the IRAM-30\,m telescope, we
observed the inner 2\,kpc of NGC\,1068 between 86.2\,GHz and 115.6\,GHz. We
identified 35 spectral features, corresponding to 24 different
molecular species. Among them, HC$_3$N, SO, N$_2$H$^+$, CH$_3$CN, NS,
$^{13}$CN, and HN$^{13}$C are detected for the first time in
NGC\,1068. Assuming local thermodynamic equilibrium (LTE), we
calculated the column densities of the detected molecules, as well as
the upper limits to the column densities of some undetected
species. The comparison among the chemistries of NGC\,1068, M\,82, and
NGC\,253, suggests that, apart from X-rays, shocks also determine the chemistry of NGC\,1068. We propose the column density ratio between CH$_3$CCH and HC$_3$N
as a prime indicator of the imprints of starburst and AGN
environments in the circumnuclear interstellar medium. This ratio is,
at least, 64 times larger in M\,82 than in NGC\,1068, and, at least, 4
times larger in NGC\,253 than in NGC,\,1068. Finally, we
used the UCL\_CHEM and UCL\_PDR chemical codes to constrain the origin of the species, as well as to test the influence of UV radiation fields and cosmic rays on the observed abundances. 

\end{abstract}

\section{Introduction}
\label{sect.Intro}

The proximity of NGC\,1068 (D=14.4\,Mpc, corresponding to a spatial
scale on 1$''=72$\,pc, \cite{Tully88}) as well as its Seyfert 2
activity turn this galaxy into the best target to study the chemical
composition of molecular clouds in the vicinity of an AGN. The
molecular material close to the central engines of AGN is thought to
be heavily pervaded by X-ray radiation originating in the nuclear
accretion disk \cite{Maloney96}. This strong radiation field has been
claimed to be responsible for the observed abundance ratios of some
species, such as HCO$^+$/HCN, or HNCO/CS, which appear to be clearly
different between AGN and starburst environments
\cite{Kohno01,Krips08,Martin09}. Observations of NGC\,1068 in a number
of other molecules, such as SiO, HOC$^+$, and CN, have revealed a
peculiar chemistry interpreted as the result of a giant X-ray
dominated region (XDR) in its nucleus, where shocks could be also
playing a role \cite{Usero04,Burillo10,Krips11}. However, an overall chemical
study of NGC\,1068 has not been done so far due to a lack of unbiased
observations of a large number of molecules. Here we present a
molecular inventory of this archetypal AGN, which is similar to those
already obtained for the starburst galaxies NGC\,253 \cite{Martin06b}
and M\,82 \cite{Aladro11b}.

\section{Observations and data analysis}
\label{sect.Obs}

\begin{figure}
\begin{center}
\includegraphics[width=0.5\textwidth]{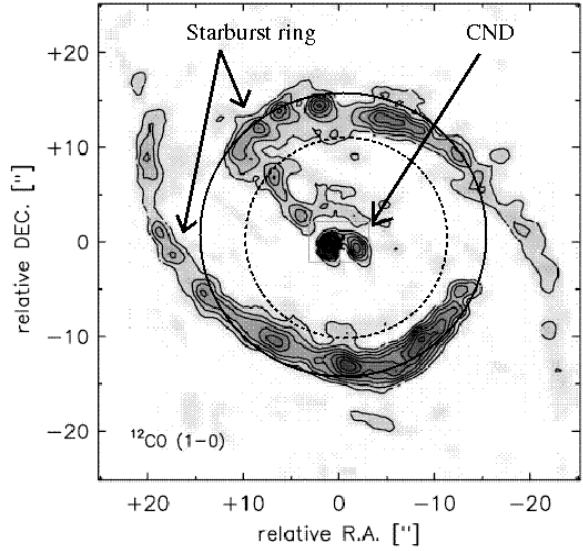}
\caption{Left: $^{12}$CO map of the central region of NGC\,1068 taken from \cite{Schinnerer00}. The beam sizes at the lowest and highest frequencies of the survey are plotted with a solid and dotted circles respectively. The starburst ring and circumnuclear disk are indicated.}
\label{12COmap}
\end{center}
\end{figure}

The observations were carried out with the IRAM-30\,m
telescope (Pico Veleta Observatory, Spain) between
October 2009 and July 2010. We observed the circumnuclear disk of NGC1068 at a
nominal position $\alpha_{2000}=$\,02:42:40.9,
$\delta_{2000}$=\,-00:00:46.0. The frequencies range observed is between
86.2\,GHz and 115.6\,GHz. We used the band E0 of the EMIR receiver and
the WILMA autocorrelator. This receiver-backend configuration allowed
us to cover 8\,GHz simultaneously in the vertical and horizontal
polarizations, and led to a channel width spacing of
$7-9$\,km\,s$^{-1}$. The data were calibrated using the standard dual load
method. The observations were done wobbling the secondary mirror with
a switching frequency of 0.5\,Hz and a beam throw of $\pm$220$''$ in
azimuth. 
We checked the pointing accuracy every hour towards the nearby bright continuum sources
2251+158 and 0113-118. The pointing corrections were always better
than 4$''$. The focus was also checked at the beginning of each run
and during sunsets. The rejection of the image sideband was better
than 10\,dB along the band. The beam sizes ranged from 21$''$ to
29$''$. This implies that the observations at the lowest frequencies
could have picked up some emission from the starburst ring, while the
highest frequencies only detected the CND emission (see
Fig.~\ref{12COmap}). We assume that, in the worse case scenario, the
starburst ring contribution is lower than 30\% \cite{Usero04}.

The observed spectra were converted from antenna temperatures ($T_{\rm A}^*$) to main beam temperatures ($T_{\rm MB}$) using the relation $T_{\rm MB}=(F_{\rm eff}/B_{\rm eff})\,T_{\rm A}^*$, where $F_{\rm eff}$ is the forward efficiency of the telescope, which values were between 0.94 and 0.95, and $B_{\rm eff}$ is the main beam efficiency, ranging from 0.77 to 0.81. Linear baselines were subtracted in all cases. The rms achieved is $\le 2$\,mK across the whole survey. The data were also corrected by the beam dilution effect as $T_{\rm B}=[(\theta^2_{\rm s}\,+\,\theta^2_{\,\rm b})\,/\,\theta^2_{\,\rm s}]\,T_{\rm MB}$, where $T_{\rm B}$ is the source averaged brightness temperature, $\theta_{\,\rm s}$ is the source size and $\theta_{\,\rm b}$ is the beam size. Based on NGC\,1068 interferometric observations of $^{12}$CO, HCN, and $^{13}$CO \cite{Helfer95,Schinnerer00}, we assumed an average source size of $4''$ for all the molecular species.

\section{Molecules in NGC\,1068} 
\label{sect.results}

We detected a total of 35 spectral features towards the AGN of
NGC1068. There are no new extragalactic molecular detections. However,
seven species or isotopologues are detected for the first time in this
galaxy, namely HC$_3$N, SO, N$_2$H$^+$, CH$_3$CN, NS, $^{13}$CN, and
HN$^{13}$C. The line identification largely follows the method
explained in \cite{Aladro11b}. We fitted Gaussian profiles to all the
detected species. The resulting Gaussian parameters, as well as
figures showing the data, and more specific comments about each
molecule, will be presented in Aladro et al in preparation.

Assuming local thermodynamic equilibrium conditions and optically thin emission, we used rotation
diagrams to determine the column densities ($N_{\rm mol}$) of the
detected species \cite{Goldsmith99}. However, some molecules, such as
CO, HCN or HCO$^+$, seem to be moderately optically thick in
NGC\,1068, as will be discussed in Aladro et al. in preparation. Therefore, their column densities are likely to be higher. On the other
hand, in those cases where we detected only one transition of a given
molecule, we assumed a rotational temperature ($T_{\rm rot}$) of
$20\pm10$\,K. The $T_{\rm rot}$ value might have an important impact
in the column density determination only in extreme cases, such as
$T_{\rm rot}<5$\,K or $T_{\rm rot}>100$\,K. Table~\ref{TableNT} lists
the $N_{\rm mol}$ obtained for each molecule, as well as the $3\sigma$
upper limits to the column densities of some other undetected
molecules in NGC\,1068.

We found that the CO species and their carbon and oxygen
isotopologues, $^{13}$CO and C$^{18}$O, are the most abundant
molecules. On the contrary, HOC$^+$ is the less abundant of the
detected species in our survey. We took from the literature some
molecular transitions that lie outside our 3\,mm frequency range (see
Table~\ref{TableNT} for references).  Few molecules show two different
components in the Boltzmann diagrams. This is the case of $^{12}$CO,
$^{13}$CO and CS. One of such components belongs to a colder gas with
higher column densities, while the other one shows warmer rotational
temperatures and lower column densities. This latter gas component is
likely to arise from the inner regions of the molecular clouds.

\begin{table}
\caption{Column densities of the observed molecules, and $3\sigma$ upper limits to the column densities of some undetected species in our survey.}

\centering

\begin{tabular}[!h]{lclclccc} 

\hline
Molecule	&	$N_{\rm mol}$	[cm$^{-2}$]	& Molecule	& $N_{\rm mol}$	[cm$^{-2}$] & Molecule	& $N_{\rm mol}$	[cm$^{-2}$]\\		
\hline
$^{12}$CO\,$^a$	&	$(4.0\pm0.3)\times10^{18}$ 	& HCO		& 	$(3.8\pm2.7)\times10^{14}$ & 	H$_2$CO		&	$\le1.1\times10^{16}$ \\
		&	$(1.2\pm0.3)\times10^{18}$ 	& SO\,$^N$	&	$(3.2\pm3.3)\times10^{14}$ & 	OCS		&	$\le5.4\times10^{14}$\\	
$^{13}$CO\,$^a$	&	$(3.3\pm0.3)\times10^{17}$ 	&  NS\,$^N$	&	$(3.2\pm1.6)\times10^{14}$ &  CH$_2$NH	&	$\le5.4\times10^{14}$	\\
		&	$(7.8\pm2.8)\times10^{16}$ 	& $^{13}$CN\,$^N$	&	$(3.2\pm1.6)\times10^{14}$ & CH$_3$CCH	&	$\le4.5\times10^{14}$	\\
C$^{18}$O\,$^b$	&	$(1.0\pm0.1)\times10^{17}$ 	& HNCO	&	$(2.8\pm0.1)\times10^{14}$ & 	SO$_2$		&	$\le3.6\times10^{14}$  	\\
C$_2$H		&	$(1.5\pm0.7)\times10^{16}$ 	& N$_2$H$^+$\,$^N$ &$(1.6\pm0.8)\times10^{14}$ & c-C$_3$H	&	$\le2.9\times10^{14}$ \\
CN		&	$(8.1\pm3.9)\times10^{15}$ 	& C$^{34}$S\,$^b$	&	$(1.6\pm0.6)\times10^{14}$ & H$_2$CS		&	$\le2.6\times10^{14}$		 \\
HCN		&	$(2.7\pm1.3)\times10^{15}$	& H$^{13}$CN&	$(1.2\pm0.6)\times10^{14}$ & C$_2$S		&	$\le8.9\times10^{13}$	\\
CH$_3$OH	&	$(2.3\pm0.1)\times10^{15}$	& H$^{13}$CN&	$(1.2\pm0.6)\times10^{14}$ & HOCO$^+$	& 	$\le7.0\times10^{13}$	\\
CS\,$^c$	&	$(7.6\pm2.2)\times10^{14}$	& SiO	&	$(9.1\pm4.5)\times10^{13}$ & $^{13}$CS	&	$\le3.6\times10^{13}$	\\
		&	$(7.0\pm4.2)\times10^{13}$	& H$^{13}$CO$^+$&	$(4.2\pm2.0)\times10^{13}$ & 	NH$_2$CN	&	$\le2.8\times10^{13}$	\\ 
HC$_3$N\,$^N$	&	$(6.0\pm4.0)\times10^{14}$	& CH$_3$CN\,$^N$	&	$(3.9\pm0.1)\times10^{13}$	 \\	
HCO$^+$\,$^e$	&	$(5.3\pm0.2)\times10^{14}$	& HN$^{13}$C\,$^N$	&$(3.2\pm5.0)\times10^{13}$	\\
HNC\,$^d$	&	$(5.2\pm0.1)\times10^{14}$	& HOC+	&	$(1.5\pm1.0)\times10^{13}$\\

\hline
\end{tabular}
\label{TableNT}

\begin{list}{}{}
\item[] $^N$:\, New detection in NGC\,1068.$^a$\,Using extra detections by \cite{Israel09}. $^b$\,Using extra detections by \cite{Martin09}. $^c$\,Using extra detections by \cite{Bayet09b}. $^d$\,Using extra detections by \cite{Perez09}. $^e$\,Using extra detections by \cite{Krips08}. To calculate the upper limits of the NGC\,1068 column densities, we assumed for all the cases a $T_{\rm rot}=20\pm10$\,K and a width of 230\,km\,s$^{-1}$. 
\end{list}{}{}
\end{table}

\section{On the origin of the molecular emission}

We used the UCL\_CHEM code \cite{Viti99,Viti04} as well as the UCL\_PDR code \cite{Bell06} to model the molecular
emission of the detected species shown in
Table~\ref{TableNT}. UCL\_CHEM is a depth- and time-dependent chemical
model that consists of two separated phases. In phase I, the free-fall
collapse of a molecular cloud with atomic composition is
simulated. The hydrogen density, initially 10$^2$\,cm$^{-3}$, is
increased until it reaches a final density (free parameter), while the
temperature remains low and constant, at 10\,K. During this phase,
atoms and molecules deplete on to the grains, and surface reactions,
including hydrogenation, take place. Phase II simulates the chemistry
of the molecular cloud {\emph{after}} a source of radiation is
switched on. We computed the temperatures at different visual extinctions using the UCL\_PDR code.
During Phase II we assume instantaneous evaporation of the molecular ices. Therefore,
only gas-phase reactions take place during this second phase. We have
considered 205 different molecules, of which 51 are surface
species. 2345 reactions, taken from the UMIST database
(http://www.udfa.net, \cite{Woodall07}), are involved in the
calculations.

Table~\ref{UCLCHEM} shows the values of the main parameters of
UCL\_CHEM. We used the NGC\,1068 metallicity (1.056,
\cite{Zaritsky94}), and the standard initial element abundance ratios
of C, O, N, S, He, Mg, and Si, compiled by \cite{Bayet08}. We run a
set of four models varying the radiation field and the cosmic-ray
ionization rate, in order to sample a variety of conditions that
may co-exist in the AGN. We used a number of assumptions when
running our models. First, we used a constant density for Phase
II. However, molecules are expected to trace environments of different
densities, probably ranging from $\sim$10$^3$\,cm$^{-3}$ to
$\sim$10$^6$\,cm$^{-3}$. Therefore, our models do not aim to simulate
density gradients within molecular clouds. Second, we took a visual
extinction of $A_{\rm v}=2$\,mag as an example of a photon-dominated
region (PDR), and a $A_{\rm v}=10$\,mag as an example of a molecular
cloud with a dense core. Third, in some cases molecular abundances
vary with time. This mainly occurs in the dense cores of the simulated
molecular clouds, where species do not reach a steady-state. We thus
averaged the abundances over a range of time from
$10^3$ to $10^7$years. However, some molecules may have shorter or larger
life cycles, and therefore their abundances may differ.

As can be seen in Table~\ref{UCLCHEM}, the four models differ in the
values of the radiation fields ($G_0=1$ and 
1000 in Harbing units) and the cosmic-ray ionization rate (
1.3$\times$10$^{-17}$\,s$^{-1}$ and
1.3$\times$10$^{-15}$\,s$^{-1}$). Model {\emph{a}} aims to simulate
the physical conditions of hot cores, which have both low external
radiation fields and cosmic ray ionization rates; in model {\emph{b}}
we increased the radiation field by a factor of 1000 with respect to
model {\emph{a}} in order to see the effects of strong UV fields in
the chemistry of the molecular clouds; on the contrary, model
{\emph{c}} has low UV field, but a high cosmic ray rate; finally,
model {\emph{d}} has high values of both parameters. Table
~\ref{Models} shows whether, according to each model, the molecules
are tracing dense gas regions, PDRs (i.e. the external zones of the molecular clouds), or both. 
 Some species seem to not be strongly affected by
the variations of UV and cosmic ray fields. Examples are C$_2$H and CN, which always trace PDRs; 
or CH$_3$OH, CS, SO, HNCO and SiO, which always trace dense gas regions. On the
other hand, other molecules such as HCO$^+$, HNC or N$_2$H$^+$, trace dense gas regions or PDRs depending on the physical conditions.

 \begin{table}
\caption{Main parameters of the UCL\_CHEM models}
\centering
\begin{tabular}[!h]{lcccccc} 
\hline
Parameter	& Model a	 &	Model b	& Model c & Model d\\
\hline
Initial H density (phase I)	& 10$^2$\,cm$^{-3}$	& 10$^2$\,cm$^{-3}$	& 10$^2$\,cm$^{-3}$	& 10$^2$\,cm$^{-3}$	\\
Final H density (phases I and II)& 10$^6$\,cm$^{-3}$ &	10$^6$\,cm$^{-3}$ & 10$^6$\,cm$^{-3}$ & 10$^6$\,cm$^{-3}$\\
Temperature (phase I)	& 10\,K & 10\,K &  10\,K & 10\,K\\
Temperature (phase II)	& 300\,K & 300\,K & 300\,K & 300\,K\\
Visual extinction	& 2 \& 10\,mag &  2 \& 10\,mag & 2 \& 10\,mag & 2 \& 10\,mag\\
Gas-to-dust mass ratio	& 100	 & 100	 & 100	 & 100\\
External UV radiation intensity	& 1 Habing & 1000 Habing & 1 Habing & 1000 Habing \\
Cosmic-ray ionization rate &	1.3$\times$10$^{-17}$\,s$^{-1}$ & 1.3$\times$10$^{-17}$\,s$^{-1}$ & 1.3$\times$10$^{-15}$\,s$^{-1}$ & 1.3$\times$10$^{-15}$\,s$^{-1}$\\
 \hline
 \end{tabular}

 \label{UCLCHEM}
 \end{table}

 \begin{table}
\caption{Regions traced by each molecule according to the four models presented in Table~\ref{UCLCHEM}.}
\centering
\begin{tabular}[!h]{lccccccccccccccc} 
\hline
Molecule	& \multicolumn{2}{|c|}{Model a}	 & \multicolumn{2}{|c|}{Model b} &  \multicolumn{2}{|c|}{Model c} & \multicolumn{2}{|c}{Model d} \\
		&    \multicolumn{1}{|c}{PDR}  &  \multicolumn{1}{|c|}{Dense Gas}	&   \multicolumn{1}{|c}{PDR}  & \multicolumn{1}{|c|}{Dense gas}	&   \multicolumn{1}{|c}{PDR}  & \multicolumn{1}{|c|}{Dense gas}	&   \multicolumn{1}{|c}{PDR}  & \multicolumn{1}{|c}{Dense gas}	\\
\hline
CO 	& \ding{51} 	&\ding{51} &  & \ding{51}& \ding{51}& \ding{51}& & \ding{51}\\
C$_2$H	& \ding{51}	& 		& \ding{51}  	& 		& \ding{51}& 		& \ding{51}	& 		\\
CN 	& \ding{51}	&       	 &\ding{51}  	&	   	& \ding{51}&	  	& \ding{51}	&		\\
HCN 	& 		& \ding{51}	&	  	& \ding{51}	& \ding{51}& \ding{51} 	& 		&  \ding{51}\\
CH$_3$OH&		&  \ding{51}	&	  	&  \ding{51}	&	    &  \ding{51}&		& \ding{51}\\
CS 	& 		& \ding{51}	&	  	&  \ding{51}    &	    & \ding{51}  &		& \ding{51}\\
HC$_3$N &		&  \ding{51}	&\ding{51}  	&  		& \ding{51}   & 	&\ding{51}	& \ding{51}\\
HCO$^+$ & \ding{51}	&	  	&\ding{51}	& \ding{51} 	&\ding{51}  &  		&  		&\ding{51}  \\
HNC 	& \ding{51}	& \ding{51}	& 	  	& \ding{51} 	&\ding{51} &\ding{51}  	&	  	&\ding{51}\\
HCO 	& \ding{51}	&	 	& \ding{51}	&       	&\ding{51}  &   \ding{51} &\ding{51}   	&       \\
SO	& 		& \ding{51}	& 	  	& \ding{51} 	& 	    & \ding{51}  & 	 	& \ding{51} \\
NS	& 		& \ding{51}	&\ding{51}  	& 		& 	    	& \ding{51}& \ding{51}  & \ding{51} \\
HNCO 	& 		& \ding{51}	&	  	& \ding{51}	& 	    	& \ding{51}& 	         & \ding{51} \\
N$_2$H$^+$& \ding{51}	& 		& 	  	&  \ding{51}	&\ding{51} 	&	  &	 	& \ding{51} \\
SiO 	& 		& \ding{51}	&	  	& \ding{51} 	&	  	&\ding{51} 	&	&\ding{51}\\
CH$_3$CN & 		& \ding{51}	&	  	&\ding{51}	& \ding{51}    	&	 &  	 	&\ding{51}\\
HOC$^+$ & \ding{51}	& \ding{51}	& \ding{51}	&\ding{51}    	& \ding{51}	&\ding{51}& \ding{51}	&\ding{51}	\\
CH$_3$CCH  &		&\ding{51}	  &	    	&\ding{51} 	&		&\ding{51} &   		&  \ding{51}\\

 \hline
 \end{tabular}

\begin{list}{}{}
\item[]Ticks show whether each molecules trace PDRs/CRDRs, dense gas, or both. Boldface ticks indicate when the abundances given by the models match the observed abundances within one order of magnitude.
\end{list}{}{}
 \label{Models}
 \end{table}

\section{Molecular abundances in AGN and starburst galaxies}
\label{comp.abund}

We aimed to compare the molecular abundances of NGC\,1068 with those of
the starburst galaxies M\,82 and NGC\,253. These are the three
extragalactic objects whose chemical compositions are better known so
far. NGC\,253 and M\,82 were also targets of unbiased molecular line
surveys in the mm range, with 25 and 18 molecular species detected in
their respective nuclear regions \cite{Martin06b,Aladro11b}. The
starburst in the center of M\,82 is old, with an average stellar
population age of $\sim15$\,Myr \cite{Konstantopoulos09}. The nuclear
starburst, together with a high supernova rate, create strong UV
fields, and thus PDRs dominate the nucleus of M\,82. On the other
hand, NGC\,253 is claimed to be in an early stage of the starburst
evolution \cite{Martin09,Aladro11a}, with younger stellar populations
in its nucleus ($\sim6$\,Myr \cite{Onti09}) where the PDRs, although
present, do not drive the heating of molecular clouds. Instead, the
nucleus of NGC\,253 is dominated by large-scale shocks between
molecular complexes. A detailed comparison of the M\,82 and NGC\,253
chemistries done by \cite{Aladro11b} showed that the ISM in both
galaxies have different composition, and that molecular abundances can
be used to distinguish the physical processes that dominate galaxy
nuclei.

\begin{figure*}
\begin{center}
\includegraphics[angle=0,width=\textwidth]{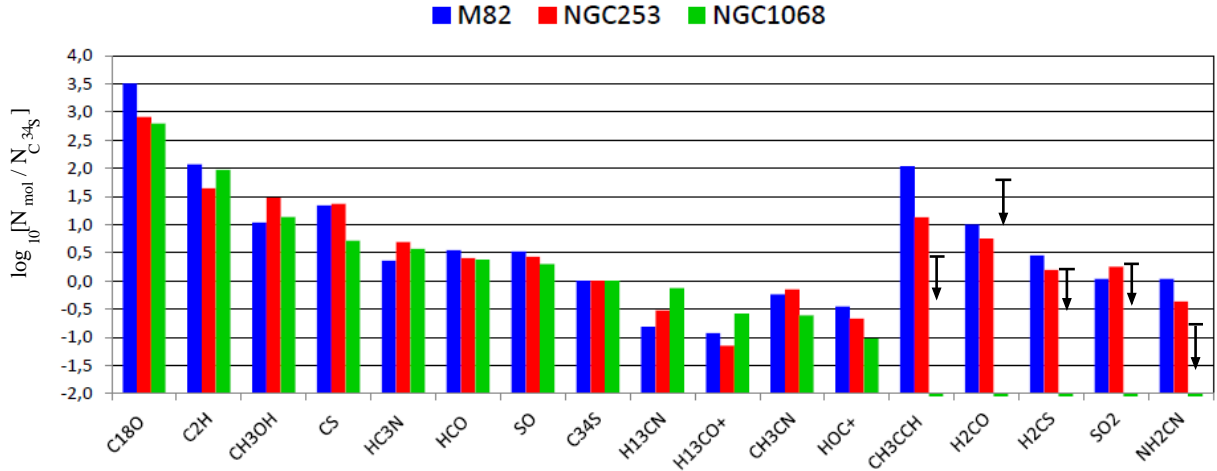}
\caption{Fractional abundances (with respect to C$^{34}$S) of several molecules in the center of the galaxies M\,82 and  NGC\,253 (data taken from \cite{Aladro11b} and references therein), and NGC\,1068. The vertical arrows show the upper limits of the fractional abundances to the undetected molecules in NGC\,1068.}
\label{Abundances}
\end{center}
\end{figure*}

Figure~\ref{Abundances} shows the molecular abundances in the central regions of M\,82, NGC\,253, and NGC\,1068. The molecular ratios were calculated with respect to C$^{34}$S because, apart from being optically thinner than the main sulfur isotopologue, it does not seem to be significantly affected by the type of nuclear activity in galaxies \cite{Martin09}. Furthermore, C$^{34}$S has similar column densities in the three galaxies (few times $10^{14}$\,cm$^{-3}$, \cite{Aladro11b}). We note that while M\,82 and NGC\,253 are located at approximately the same distance, $\sim3$\,Mpc \cite{Freedman94}, NGC\,1068 lies nearly five times further, at 14.4\,Mpc \cite{Tully88}. While the comparison of the  column densities in these galaxies would be biased by the distance, this should not affect the comparison of the fractional abundances. On the other hand, because of the different distances the IRAM-30m beam covered the $1.5-2.0$ central kiloparsecs of the AGN, while for the starburst galaxies, the beam covered the inner 360\,pc.

Among all the species compared, only H$^{13}$CN and H$^{13}$CO$^+$
show higher abundances with respect to C$^{34}$S in NGC\,1068 than M\,82 and NGC\,253. On the other hand, C$_2$H, CH$_3$OH, HC$_3$N, HCO, and SO show similar
abundances in all three galaxies (within a factor three). A third
group is composed by those molecules that are clearly more
abundant in M\,82 and NGC\,253 than in NGC\,1068. From this group it
can be drawn the clearest chemical differentiation between starbursts
and AGN activities. Among the molecules detected in the three
galaxies, CS/C$^{34}$S is almost 5 times higher in NGC\,253 than in
NGC\,1068. Similarly, CH$_3$CN/C$^{34}$S shows a difference of a
factor 3 between NGC\,253 and NGC\,1068, while HOC$^+$/C$^{34}$S is a
factor 4 more abundant in M\,82 than in NGC\,1068. It is remarkable
that these three molecular ratios are very similar in M\,82 and
NGC\,253 (see Fig.~\ref{Abundances}). Moreover, some of the biggest
chemical differences are given by those molecules that were clearly
detected in both starburst galaxies, but not in NGC\,1068, such as
CH$_3$CCH, H$_2$CO, H$_2$CS, SO$_2$ and NH$_2$CN. Among
them, CH$_3$CCH/C$^{34}$S is the ratio showing the highest contrast,
being $>39$ times higher in M\,82 than in NGC\,1068 and $>5$ times
higher in NGC\,253 than in NGC\,1068. Therefore, we computed other
fractional abundances of CH$_3$CCH with respect to the molecules
having similar abundances in the three galaxies. We found that the
most striking differences among the AGN and the starburst galaxies are given
by R$_{\rm CH_3CCH/HC_3N}$ and R$_{\rm CH_3CCH/C_2H}$. CH$_3$CCH and
HC$_3$N are claimed to be tracers of dense gas in galaxies
\cite{Aladro11a}, while C$_2$H might be tracing
both PDRs and dense gas regions
\cite{Bayet09a}. To avoid comparisons of
molecules that might be arising from different regions, we select
[CH$_3$CCH/HC$_3$N] as the best ratio to chemically differentiate
between AGN and starburst galaxies. We found that R$_{\rm
CH_3CCH/HC_3N}>64$ when comparing M\,82 with NGC\,1068, and R$_{\rm
CH_3CCH/HC_3N}>4$ when comparing NGC\,253 with NGC\,1068. We note that
a factor of, at least, 64 is much larger than the R$_{\rm HCO^+/HCN}$
commonly used to differentiate between AGN and starburst environments
\cite{Kohno01, Krips08}. On the other hand, our upper limit to the
column density of CH$_3$CCH in NGC\,1068 is not very tight, so the
molecular ratios could be even larger. The values found for R$_{\rm
CH_3CCH/HC_3N}$ indicate that the dense ISM in the NGC\,1068 nucleus
resembles more NGC\,253 than M\,82. This points at the existence of
shocks (rather than PDRs) accompanying the strong X-ray fields present
in the central regions of NGC\,1068.

\section{Conclusions}
\label{conclusions}

We carried out a molecular line survey with the IRAM 30-m telescope
towards the active galactic nucleus of NGC\,1068. We covered the
frequencies between 86.2\,GHz and 115.6\,GHz, detecting a total of 35
spectral line features. Seven species, or their carbon isotopologues,
are detected for the first time in this galaxy, namely HC$_3$N, SO,
N$_2$H$^+$, CH$_3$CN, NS, $^{13}$CN, and HN$^{13}$C.

Some molecules, such as CH$_3$CCH, c-C$_3$H$_2$, OCS, SO$_2$, NS,
HOCO$^+$, and NH$_2$CN, are not detected in NGC\,1068. We compared the
molecular abundances of the NGC\,1068 nucleus (claimed to be a giant
XDR, e.g. \cite{Burillo10}), with those of M,82 (claimed to be a giant
PDR, \cite{Fuente05}), and NGC\,253 (claimed to be dominated by
low-velocity shocks, \cite{Martin06b}). We found that the ratio
R$_{\rm CH_3CCH/HC_3N}$ shows the largest differences among the three
galaxies. Specifically, R$_{\rm CH_3CCH/HC_3N}$ is at least 64 times
larger in M\,82 than in NGC\,1068, and at least 4 times larger in
NGC\,253 than in NGC\,1068. Therefore, this ratio can be used to
disentangle between AGN and starburst physical
environments. Furthermore, R$_{\rm CH_3CCH/HC_3N}$ indicates that UV
fields are not probably strong enough in the inner 2\,kpc of
NGC\,1068. Instead, shocks, could be playing a secondary role, after
X-rays.

We used the time-dependent chemical (UCL\_CHEM) and PDR (UCL\_PDR) models 
to simulate the
chemistry in molecular clouds under several physical conditions that
might co-exist in AGN. We note, however, that our aim was not to
simulate the particular environment in the center of NGC\,1068, but to
study the influence of UV and cosmic ray fields in the abundances of a large number of molecules. We found
that some species seem to always trace photon-dominated regions
and/or cosmic ray-dominated regions independently of the strength of
the UV and cosmic ray fields (such as C$_2$H and CN), while
other species always trace dense gas regions (e.g. CH$_3$OH, CS,
SO, HNCO, and SiO). Finally, some molecules might be
arising from the inner and/or outer layers of the molecular clouds depending on the strength of the external radiation fields (examples
are HCO$^+$, HNC and N$_2$H$^+$).


\section*{References}

\begin{thebibliography}{200} 

\bibitem{Aladro11a} Aladro, R., Mart{\'{\i}}n-Pintado, J., Mart{\'{\i}}n, S., Mauersberger, R., \& Bayet, E.\ 2011a, A\&A, 525, A89 

\bibitem{Aladro11b} Aladro, R., Mart{\'{\i}}n, S., Mart{\'{\i}}n-Pintado, J., et al.\ 2011b, A\&A, 535, A84 

\bibitem{Bayet08} Bayet, E., Viti, S., Williams, D.~A., \& Rawlings, J.~M.~C.\ 2008, ApJ, 676, 978 

\bibitem{Bayet09a} Bayet, E., Viti, S., Williams, D.~A., Rawlings, J.~M.~C., \& Bell, T.\ 2009a, ApJ, 696, 1466 

\bibitem{Bayet09b} Bayet, E., Aladro, R., Mart{\'{\i}}n, S., Viti, S., \& Mart{\'{\i}}n-Pintado, J.\ 2009b, ApJ, 707, 126 

\bibitem{Bell06} Bell, T.~A., Roueff, E., Viti, S., \& Williams, D.~A., 2006, 
MNRAS, 371, 1865 

\bibitem{Onti09} Fern{\'a}ndez-Ontiveros, J.~A., Prieto, M.~A., \& Acosta-Pulido, J.~A., 2009, MNRAS, 392, L16 

\bibitem{Freedman94} Freedman, W.~L., et al.\ 1994, ApJ, 427, 628 

\bibitem{Fuente05} Fuente, A., Garc{\'{\i}}a-Burillo, S., Gerin, M., et al.\ 2005, ApJL, 619, L155 

\bibitem{Burillo10} Garc{\'{\i}}a-Burillo, S., et al.\ 2010, A\&A, 519, A2 

\bibitem{Goldsmith99} Goldsmith, P.~F., \& Langer, W.~D.\ 1999, ApJ, 517, 209 

\bibitem{Helfer95} Helfer, T.~T., \& Blitz, L.\ 1995, ApJ, 450, 90 

\bibitem{Israel09} Israel, F.~P.\ 2009, A\&A, 493, 525 

\bibitem{Kohno01} Kohno, K., Matsushita, S., Vila-Vilar{\'o}, B., et al.\ 2001, The Central Kiloparsec of Starbursts and AGN: The La Palma Connection, 249, 672 

\bibitem{Konstantopoulos09} Konstantopoulos, I.~S., Bastian, N., Smith, L.~J., Westmoquette, M.~S., Trancho, G., \& Gallagher, J.~S.\ 2009, ApJ, 701, 1015 

\bibitem{Krips08} Krips, M., Neri, R., Garc{\'{\i}}a-Burillo, S., et al.\ 2008, ApJ, 677, 262 

\bibitem{Krips11} Krips, M., Mart{\'{\i}}n, S., Eckart, A., et al.\ 2011, ApJ, 736, 37 

\bibitem{Maloney96} Maloney, P.~R., Hollenbach, D.~J., \& Tielens, A.~G.~G.~M.\ 1996, ApJ, 466, 561 

\bibitem{Martin06b} Mart{\'{\i}}n, S., Mart{\'{\i}}n-Pintado, J., \& Mauersberger, R.\ 2006, A\&A, 450, L13      

\bibitem{Martin09} Mart{\'{\i}}n, S., Mart{\'{\i}}n-Pintado, J., \& Mauersberger, R.\ 2009, ApJ, 694, 610 

\bibitem{Myers87} Myers, S.~T., \& Scoville, N.~Z.\ 1987, ApJl, 312, L39 

\bibitem{Perez09} P{\'e}rez-Beaupuits, J.~P., Spaans, M., van der Tak, F.~F.~S., et al.\ 2009, A\&A, 503, 459 

\bibitem{Schinnerer00} Schinnerer, E., Eckart, A., Tacconi, L.~J., Genzel, R., \& Downes, D.\ 2000, ApJ, 533, 850 

\bibitem{Tully88} Tully, R.~B.\ 1988, Nearby Galaxies Catalog, Cambridge University Press

\bibitem{Usero04} Usero, A., Garc{\'{\i}}a-Burillo, S., Fuente, A., Mart{\'{\i}}n-Pintado, J., \& Rodr{\'{\i}}guez-Fern{\'a}ndez, N.~J.\ 2004, A\&A, 419, 897 

\bibitem{Viti99} Viti, S., \& Williams, D.~A.\ 1999, MNRAS, 305, 755 

\bibitem{Viti04} Viti, S., Collings, M.~P., Dever, J.~W., McCoustra, M.~R.~S., \& Williams, D.~A.\ 2004, MNRAS, 354, 1141 

\bibitem{Woodall07} Woodall, J., Ag{\'u}ndez, M., Markwick-Kemper, A.~J., \& Millar, T.~J.\ 2007, A\&A, 466, 1197 

\bibitem{Zaritsky94} Zaritsky, D., Kennicutt, R.~C., Jr., \& Huchra, J.~P.\ 1994, ApJ, 420, 87 


\end{thebibliography}
\end{document}